\documentstyle[11pt]{article}

\input psfig.sty

\def\Journal#1#2#3#4{{#1} {\bf #2}, #3 (#4)}

\def\NIMA{{\em Nucl. Instrum. Methods} A}
\def\NPB{{\em Nucl. Phys.} B}
\def\PLB{{\em Phys. Lett.}  B}
\def\PLA{{\em Phys. Lett.}  A}
\def\PRL{\em Phys. Rev. Lett.}
\def\PRD{{\em Phys. Rev.} D}
\def\PRA{{\em Phys. Rev.} A}
\def\JPA{{\em J. Phys.} A}
\def\ZPC{{\em Z. Phys.} C}
\def\PA{\em Particle Accelerators}

\def\NAT{\em Nature}
\def\CMP{\em Comm. Math. Phys.}
\def\PNAS{\em Proc. Nat. Acad. Sci.}
\def\DAN{\em Dokl. Akad. Nauk. SSR)}
\def\SPD{\em Sov. Phys. Dokl.}
\def\JETP{\em Sov. Phys.-JETP}
\def\ZETF{\em Zh. Eksp. Teor. Fiz.}
\def\UFN{\em Usp. Fiz. Nauk.}
\def\SPU{\em Sov. Phys.-Usp.}
\def\RMP{\em Rev. Mod. Phys.}

\def\be{\begin{equation}}
\def\ee{\end{equation}}
\def\bea{\begin{eqnarray}}
\def\eea{\end{eqnarray}}

\newcommand \half {\frac{1}{2}}
\newcommand\ie {{\it i.e. }}

\begin{document}

\vspace* {-35 mm}
\begin{flushright}  OSLO-TP 3-98\\
April-1998
\end{flushright}

\vskip 10mm
\centerline{\Large\bf Accelerated Electrons and the Unruh Effect 
\footnote{To appear in the Proceedings of the Advanced ICFA Beam Dynamics
Workshop on Quantum Aspects of Beam Physics, Monterey January 4-9, 1998, ed.
Pisin Chen.}} 

\vskip 10mm
\centerline{Jon Magne Leinaas}               
\begin{center}
{\em Department of Physics \\
P.O. Box 1048 Blindern \\
N-0316 Oslo \\
Norway}
\end{center}
\vskip 10mm
\newcommand \sss {\mbox{ $<\overline{s}s>$} }
\def\fk{\mbox{ $f_K$} }
\centerline{\bf Abstract}
\vskip 3mm

\noindent
Quantum effects for electrons in a storage ring are
studied in a co-moving, accelerated frame. The polarization effect due to
spin flip synchrotron radiation is examined by treating the electron as a
simple quantum mechanical two-level system coupled to the orbital motion and
to the radiation field. The excitations of the spin system are related to the
Unruh effect, \ie the effect that an accelerated radiation detector is
thermally excited by vacuum fluctuations. The importance of orbital
fluctuations is pointed out and the vertical fluctuations are examined.

\section{Introduction}
The meeting at Monterey has been dedicated to the memory of John Bell, and I
would like to dedicate also the present paper to his memory. The paper is
based on a work which I did with him several years ago \cite{Bell83,Bell87}
and on a talk I gave at CERN in 1991 \footnote{I am indebted to Pisin Chen
for his invitation to include this paper in the proceedings from the Monterey
workshop. A written version of the talk at CERN has been planned to be
published, but has been long delayed.}. 

The collaboration with John Bell started when, as a fellow at CERN, I
discussed with him the rather exotic theoretical effect often referred to as
the Unruh effect. As shown by Davies \cite{Davies75} and Unruh
\cite{Unruh76}, an idealized radiation detector which is accelerated through
ordinary Minkowski vacuum gets heated due to interactions with the vacuum
fluctuations of the radiation field. For uniform linear acceleration the
excitation spectrum has a universal, thermal form, independent of details of
the detector. This effect, that vacuum seems hot, as measured in an
accelerated system, has by Unruh and others been related to the phenomenon
of Hawking radiation from black holes
\cite{Hawking74}.

The idea came up during our discussion that a depolarization effect which
was known to exist for electrons in a storage ring could have something to do
with the Unruh heating. We investigated this and found that the effect
indeed was related, although there were important complications due to the
fact that the electrons were following a circular orbit rather than being
linearly accelerated \cite{Bell83}. Motivated by these complications we later
examined more carefully the effects of quantum fluctuations for an electron
moving in a circular orbit \cite{Bell87}.

In this paper I review the description of the polarization effects for
circulating electrons, in the way discussed in our two papers. The spin
excitations are studied in a co-moving, accelerated frame, which follows the
classical path of the circulating electron.  I first discuss the description
of spin motion in the accelerated frame and then examine spin excitations
produced either directly through fluctuations in the magnetic field or
indirectly through fluctuations in the path. The close connection to the
Unruh effect for a linearly accelerated system is demonstrated, and the
polarization effect is compared to a simple two-level model with thermal
excitation of the upper level. The vertical fluctuations in the path are
shown to give an independent demonstration of the circular Unruh effect.     

\section{Quantum effects for accelerated electrons}

The motion of particles in accelerators can mostly be understood and
described in classical terms. But there are some quantum effects which are
non-negligible and which even may be important. These mainly have to do with
radiation phenomena and with the radiation reaction on the accelerated
particles. Therefore they are much more important for the light electron
than for the much heavier proton.

The accelerated electrons emit radiation, synchrotron radiation, as it is
known for particles in a magnetic field. Even for high energy electrons this
process is well described by the classical radiation formula. This was
explicitly demonstrated by Schwinger \cite {Schwinger54} who calculated the
lowest order quantum correction to the radiated power. Only for extremely
high energetic electrons the quantum corrections become important. The
condition for this being small can be written as
\be
\gamma \ll \gamma_c 
\ee
with 
\be
\gamma =
\frac{1}{\sqrt{1-(\frac{v}{c})^2)}}, \; \;\;\;
\gamma_c=\sqrt{\frac{mc\rho}{\hbar}} 
\ee
where $m$ is the electron mass, $v$ its velocity and $\rho$ is the
radius of curvature of the particle orbit. The important ratio then is 
\be
\Upsilon = (\frac{\gamma}{\gamma_c})^2 \approx \frac{a}{a_m}, \; (\gamma \gg 1)
\ee
where $a$ is the acceleration of the particle in an inertial rest frame,
and $a_m$ is an acceleration parameter determined by the particle mass,
\be
a = \frac{\gamma^2 v^2}{\rho}, \\\ a_m = \frac{mc^3}{\hbar}
\ee
Thus, the important physical quantity is the acceleration $a$ rather than
the energy of the electron. A typical value for the parameter $\Upsilon$ in
cyclic accelerators is $10^{-6}$, which shows that quantum effects indeed
are very small.\footnote{There have been some studies also of the
quantum regime $\Upsilon>1$ in connection with linear accelerators at extremely
high energies \cite{Himel85,Noble87,Jacob87,Bell88}.}

Even if synchrotron radiation is essentially a classical phenomenon for
$\gamma \ll \gamma_c$, this does not mean that all quantum effects
associated with this phenomenon are unimportant. The radiation reaction will
excite orbital oscillations even for much smaller energies \cite{Sokolov71}.
The radiation field then acts in two ways on the particle. Quantum
fluctuations excite the oscillations, while radiation damping tends to
reduce the oscillation amplitudes. Balance between these two tendencies
defines a minimal, quantum limit to the beam size.

A perhaps even clearer demonstration of quantum effects for electrons
in a storage ring is associated with the phenomenon of spin-flip radiation
\cite{Sokolov63,Derbenev73}. The asymmetry between up and down flips in
the magnetic field leads to a gradual build up of transverse
polarization of the electrons. Under ideal conditions the polarization
approaches equilibrium as \be
P(t) = P_0(1-e^{-t/t_0}),
\ee
with a maximum polarization
\be
P_0=\frac{8}{5\sqrt{3}} = 0.924
\ee
and a build up time 
\be
t_0 = \frac{8}{5\sqrt{3}} \frac{m^2c^2\rho ^3}{e^2\hbar \gamma ^5}
\ee
For existing accelerators this build up time is of the order of minutes to
hours.

The phenomenon of spontaneous polarization of electrons circulating in a
magnetic field has been analyzed in many publications, both for ideal
conditions and for the more realistic situation with particles moving in a
variable magnetic field. There also exist review articles on this interesting
subject \cite{Baier71,Jackson76,Montague84}. In particular the paper by J.D.
Jackson focuses attention on the aspects of this phenomenon which can be
described in elementary terms. My approach here is along the same line. But
whereas Jackson rejects the idea of a simple description of the effect as a
transition between spin energy levels caused by radiation effects, which then
would lead eventually to all particles in the lowest energy level, this is in
fact the picture I will use. The electron spin is treated as two-level
quantum system interacting with the radiation field. But the effect of the
radiation field along the accelerated orbit of the electrons is different from
the effect on an electron sitting at rest. Transitions also to the upper
energy level are induced by the field along this orbit, and that leads to a
small, but non-vanishing depolarization of the electron beam. I will consider
the ideal case of electrons moving in a rotationally invariant magnetic field.
A stable, classical circular orbit is produced by a radial gradient in the
magnetic field. This corresponds to the situation of a weak focussing machine.

\section{Hamiltonian for the accelerated frame}

When we introduce a co-moving frame for the description of the spin motion of
the electron, it is of interest to note that this frame is uniquely defined
only up to a (time-dependent) rotation. There are in fact three different
co-moving frames which are characterized by simple properties, in one way or
the other. The first one, which I denote the L-frame is the one which is
obtained from the fixed lab frame by a pure boost. This frame is
non-rotational as seen from the lab. The other one is the frame which rotates
with the frequency of the orbiting particle. In this frame, which I denote
the O-frame, the accelerating field is stationary. Finally there is a frame,
denoted the C-frame, which is non-rotational when viewed along the particle
orbit. The fact that this is different from the L-frame is a relativistic
effect and gives rise to Thomas precession of the spin vector.

The relative rotational frequencies of these three frames are listed in Table 1
for an electron following a circular path in a magnetic field. Both the
magnetic field $B$ and the frequencies refer to co-moving frames.
\begin{table}[t]
\caption{{\em Rotation frequencies of three different co-moving frames and
the spin precession frequencies in these frames for an electron orbiting in a
constant magnetic field  $B$}}
\vspace{0.2cm}
\begin{center}
\footnotesize
\begin{tabular}{|c|c|c|}
\hline
$frames$ & $rotational\; freq.$ & $spin \; precession$ \\ [2mm]
\hline
$C$&0& $g\frac{e}{2mc}B$   \\[2mm]
$O$&$\frac{e}{mc}B$&$(g-2)\frac{e}{2mc}B$  \\[2mm]
$L$&$(1-\frac{1}{\gamma})\frac{e}{mc}B$&$((g-2)+\frac{2}{\gamma})\frac{e}{2mc}B$
\\[2mm]
\hline
\end{tabular}
\end{center}
\end{table}
Also the spin precession frequencies in the three different frames are
shown. This has the simplest in the C-frame. Since this frame is
non-rotational along the orbit, all spin precession in this frame is due to
coupling between the magnetic moment of the particle and the external
magnetic field. Thus, the precession frequency is proportional to the
gyromagnetic factor $g$, as shown in the table. In the O-frame on the other
hand the precession frequency is proportional to $g-2$. This demonstrates the
well-known fact that for $g=2$ the spin precesses exactly with the frequency
of the orbital motion. Finally, in the L-frame there is a further correction
due to the relative rotation of the L- and the O-frame. This correction is
identical to the rotation frequency of the orbital motion, and this is
smaller than the frequency associated with the Thomas precession by a factor
of $1/\gamma$.

Even if the spin motion is simplest in the C-frame, I shall in the following
apply the O-frame for the quantum description. The reason for this is that
the external fields are stationary in this frame and therefore give rise to
a time-independent Hamiltonian. This frame will be extended to a local
accelerated coordinate system to allow for fluctuations in the particle
about the circular path, which then is assumed to be the classical, stable
orbit of the electrons. This coordinate system will necessary contain
coordinate singularities at some distance from the orbit. But I will assume
fluctuations away from the stable orbit to be small, so that linearized
equations are sufficient. I will also assume velocities in this frame to be
small, so that a non-relativistic approximation can be applied.

The Hamiltonian, which governs the time evolution in the accelerated frame
is not identical to that of the inertial rest frame, but it can be expressed
in a simple way in terms of observables from this frame,
\be
H'=H-\frac{a}{v} J_z+\frac{a}{c} K_x
\ee
Here $H$ is the Dirac Hamiltonian, $J_z$ is the generator of rotations in
the plane of the electron orbit and $K_x$ is a boost operator. The
coordinates in the O-frame then are chosen with the particle acceleration
in the negative $x$-direction and with the orbit velocity in the positive
$y$-direction. The additional terms in the expression for $H'$ are fairly
easy to understand. The presence of the generator of rotations is related
to the fact that the coordinate axes of the O-frame rotate along the orbit
and the presence of the boost operator accounts for the continuous jumping
between inertial frames when the particle is accelerated.
For the three operators included in $H'$ we have the following expressions,
\be
H=c\vec {\alpha} \cdot \vec{\pi} + \beta mc^2
+e\phi+\kappa\frac{e\hbar}{2mc}(i\beta \vec{\alpha} \cdot \vec{E}-\beta
\vec{\sigma} \cdot \vec{B}) 
\ee
\be
J_z=(\vec{r}\times \vec{p})_z + \half \hbar \sigma _z
\ee
\be
K_x=-\frac{1}{2c} (xH +Hx)
\ee
where a term for the anomalous magnetic moment, $\kappa =\half (g-2)$, has
been introduced in the expression for $H$. All the potentials and fields in
these expressions refer to the inertial rest frame of the classical orbit.
The notation $\vec{\pi} = \vec{p}-\frac{e}{c} \vec{A}$ has been used for the
mechanical moment of the electron.

When the fluctuations around the classically stable orbit are assumed to
be small, then a non-relativistic approximation makes sense. A
Foldy-Wouthuysen transformation, where we keep only the leading terms,
gives a Hamiltonian which then can be split into a spin-independent and a
spin dependent term in the following way,
\be
H = H_{orb} + H_{spin}
\ee
\be
H_{orb} = (1-\frac{ax}{c^2})(\frac{\vec{\pi}^2}{2m} + e\phi) - axm +
\frac{ia\hbar}{2mc^2} \pi_x - \frac{a}{v} (\vec{r} \times \vec{p})_z + ...
\ee
\be
H_{spin} =-\frac{e\hbar}{4mc} \vec{\sigma} \cdot [(1-\frac{ax}{c^2}) g
\vec{B} + \frac{g-1}{mc} \vec{E} \times \vec{\pi}] - \half
\frac{a\hbar}{v} \sigma _z + \frac{a\hbar}{4mc^2} (\vec{\sigma} \times
\vec{\pi})_x + ...
\ee

The Hamiltonian in the accelerated frame includes some complications
compared to that of the inertial frame. However, if we consider the
orbital motion only to linear order in the deviation from the stable
orbit, the main difference in the expressions for $H_{orb}$ is the
presence of the centrifugal and Coriolis terms. For the orbital motion the
spin effects only give rise to a small perturbation. But also for the
spin motion the effect of the fluctuations in the orbit is small, since the
spin precession mainly is determined by the strong magnetic field along the
classical orbit. In principle one could then determine the particle motion in
the following way: One first solves for the orbital motion, neglecting the
spin, and then one determines the spin motion, treating the orbital
fluctuations and the quantum fields as perturbations. However, when
calculating the polarization of the particle beam, one can simplify this
approach somewhat, since it is only the fluctuations in the particle orbit
which are driven by the coupling to the radiation field which are important. 

\section{Spin transitions}
The spin Hamiltonian can be written in the form
\be
H_{spin}=\half \hbar \vec{\omega} \cdot \vec{\sigma}
\ee
\be
\vec{\omega}=\vec{\omega} _0 + \delta \vec{\omega}
\ee
with $\vec{\omega}_0$ giving rise to the classical part of the precession,
\be
\vec{\omega}_0=-\frac{e}{2mc}g\vec{B}_0 -\frac{a}{v} \vec{k}
            = - \frac{e}{2mc}(g-2) B_0 \vec{k}
\ee and $\delta \vec{\omega}$ as the fluctuation part,
\be
\delta \vec{\omega}= -\frac{e}{2mc}[g \delta \vec{B} -
\frac{ax}{c^2}g\vec{B}_0 - (g-2) \frac{a}{ec} \vec{i} \times \vec{\pi}]
\ee
In the last two expressions $\vec{k}$ is the unit vector orthogonal to the
plane of motion and $\vec{i}$ a unit vector in the $x$-direction.
$\vec{B}_0=B_0 \vec{k}$ is the external magnetic field on the classical
orbit, and $\delta \vec{B}$ accounts for the fluctuations in the magnetic
field. This can be separated into two parts,
\be
\delta \vec{B} = \vec{B}_q + \delta \vec{B}_{c}
\ee
where $\vec{B}_q$ denotes the quantum field along the classical orbit
and $\delta \vec{B}_{c}$ is the variation in the external field due
to fluctuations in the orbit.

The spin motion now can be determined by time dependent perturbation
theory. $\vec{\omega}_0$ then defines the unperturbed part of the spin
Hamiltonian and $\delta \vec{\omega}$ the perturbation. To first order,
the transition probabilities per unit time between the levels of the
unperturbed Hamiltonian are given by
\bea
\Gamma_{\pm} &=& \lim_{T\rightarrow \infty} \frac{1}{4T} 
|\int_{-T/2}^{T/2} e^{\pm i\omega_0 \tau} \delta \omega _{\mp}
(\tau)|0>|^2 \nonumber \\ 
&=& \frac{1}{4}\int_{-\infty}^{+\infty}d\tau e^{\mp i\omega_0 \tau}<0|\delta
\omega_{\pm}(\tau/2) \delta \omega_{\mp}(-\tau/2)|0>
\eea
$|0>$ in this equation denotes the state of the combined system of radiation
field and orbit variables, unperturbed by the spin. $\delta \omega
_\pm$ is a linear combination of the $x$- and $y$-component of $\delta
\vec{\omega}$, \be
\delta \omega _\pm = \delta \omega_x \pm i \delta \omega _y  \nonumber
\ee
The same notation will be used for other variables in the following.

A useful substitution rule which can be used in the expression for
$\delta \omega _\pm$ is the following one,
\be
\frac{d}{d\tau} F \rightarrow \pm i\omega_0 F  \label{subs}
\ee
The difference between these two expressions only gives rise to end
effects in the integral for the transition amplitude, and for large $T$
this is suppressed in $\Gamma_\pm$ due to the prefactor $1/T$.
This substitution rule now can be used to eliminate the orbital variables in
the expression for $\delta \omega_\pm$, which can be written in the form
\be
\delta \omega_\pm = - \frac{e}{2mc}[gB_{q\pm} + g \delta B_{c\pm}
\pm2iv\omega_0 \frac{m}{ec}\dot{z}]    \label{deltaomega}
\ee
With the stable orbit in the symmetry plane of the magnetic field, $\delta
B_{c\pm}$ gets contribution only from the gradient in the z-direction.
This implies that (to lowest order) $\delta \omega_\pm$ only depends on the
vertical fluctuations in the particle orbit. These fluctuations in turn are
determined by coupling to the radiation field in the following way, \be
\ddot{z}-\frac{2e^2}{3mc^3}(\stackrel{...}{z}-\frac{a^2}{c^2}
\dot{z})+\Omega^2 z =\frac{e}{m} E_{qz} \ee where a radiation reaction term
has been introduced and where non-linear terms have been neglected. The
restoring electric force in the z direction can be related to the gradient in
the magnetic field, \be
\Omega^2 = \frac{a}{\rho} n
\ee
\be
n = \frac{\rho}{B_0} \frac{\partial B_z}{\partial x} = \frac{\rho}{B_0} 
\frac{\partial B_x}{\partial z}
\ee
$\rho$ is the radius of the (classical) electron orbit, and $n$ the fall-off
parameter of the magnetic field.

By use of the substitution rule eq.(\ref{subs}) now can be solved for
z, \be
z = [\Omega^2 - \omega_0^2 \pm i\Delta \omega_0]^{-1} \frac{e}{m} E_{qz} +
\Lambda
\ee
\be
\Delta = \frac{2e^2}{3mc^3} (\frac{a^2}{c^2} + \omega_0^2)
\ee
$\Lambda$ here denotes a term which is suppressed for large $T$. When the 
expression for $z$ is inserted in eq.(\ref{deltaomega}),
this gives (for $v\approx c$),
\be
\delta \omega_\pm = - \frac{e}{2mc}[gB_{q\pm} + (2+f_\pm(g)) E_{qz}] 
\label{deltaomega2}
\ee
with $f_\pm (g)$ as a resonance term,
\be
f_\pm(g) = \frac{(g-2)\Omega^2}{\Omega^2-\omega_0^2\pm i\Delta \omega_0}
\ee
This term blows up when the frequency of the free oscillations in the
z-direction is close to the classical spin precession frequency, but it tends
rapidly to zero away from the resonance.

The new expression for $\delta \omega_\pm$ (\ref{deltaomega2})
 now only depends on the free quantum fields, and the transition
probabilities can be expressed in terms of correlation functions of
these fields along the particle orbit,
\be
\Gamma_\pm = \frac{1}{4} \int_{-\infty}^{+\infty}d\tau e^{\mp
i\omega_0\tau} <0|\delta\omega_\mp(\tau/2)\delta\omega_\pm(-\tau/2)|0> 
\label{gamma}
\ee
Now $|0>$ refers to the vacuum
state of the radiation field. To calculate these probabilities now is 
straightforward. The fields in the co-moving frame most conveniently are
expressed in terms of lab frame fields and the correlation functions of
these are found by expressing the field operators in terms of creation and
annihilation operators. To leading order in $1/\gamma$ the relevant
fourier integrals can be calculated analytically. 
\begin{figure}[t]
\centerline{\psfig{figure=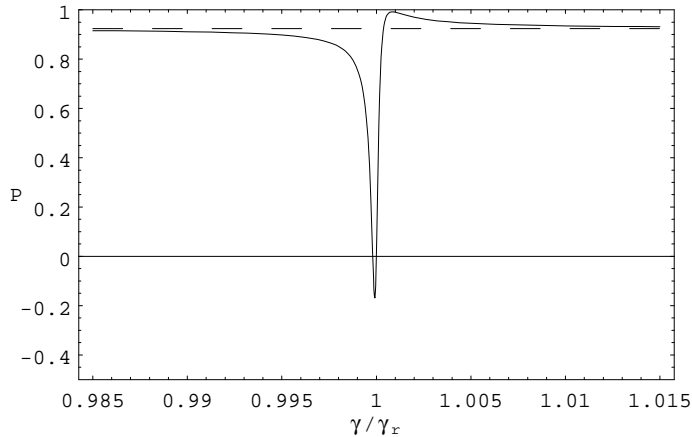,height=2in}}
\vspace{15pt}
\caption{{\em The equilibrium polarization $P$ as a function of $\gamma$
close to the resonance with the vertical oscillations. The dashed
line corresponds to the limiting value $P=0.92$, away from the
resonance. The scale for $\gamma$ is relative to the resonance value
$\gamma_r$.}  \label{fig1}}
\end{figure}
The polarization is determined by the
population of the two spin levels, and this in turn is found by the
standard  argument of equilibrium between transitions up and down. We
have, 
\be 
P=\frac{\Gamma_+ - \Gamma_-}{\Gamma_+ + \Gamma_-} 
\ee 
In Fig.\,1 the polarization is shown as a function of $\gamma$. Except for
values close to the resonance with the vertical motion, the standard result
for the polarization is found, $P = 0.924$. The effect of the resonance is
mainly to depolarize the beam, but an interesting detail is the coherent
effect which gives a maximum value of $P=0.992$ close to the resonance.
Thus, at least in principle it is possible to exceed the limiting  value of
0.924.

\section{Quantum fluctuations and the Unruh effect}
The expression for the transition probabilities $\Gamma_\pm$ now makes it
possible to see the close relation between the polarization effect and
the Unruh effect \cite{Unruh76}. Let me rewrite it in the form
\be
\Gamma_\pm = \int_{-\infty}^{+\infty} e^{i\omega_0\tau}C_\pm(\tau)
\label{gamma2}
\ee
with
\be
C_+(\tau) = <D^\dagger(0) D(\tau)> \label{C+}
\ee
\be
C_-(\tau) = <D(\tau) D^\dagger(0)> \label{C-} 
\ee
I have here introduced the new notation $D=(1/2)\delta \omega_+,
D^\dagger=(1/2)\delta \omega_-$. The operator $D$ then is a linear
combination of electric and magnetic fields in the co-moving frame,
\be
D(\tau) = \vec{\alpha} \cdot \vec{E}(x(\tau)) + \vec{\beta} \cdot
\vec{B}(x(\tau)), \label{D(tau)}
\ee
where $x(\tau)$ is the space-time orbit of the particle. The expression
(\ref{gamma2}) is similar to that which defines transitions in a point
detector in the case of the Unruh effect. The main difference is that the
world line of the detector then corresponds to linear acceleration rather
than to circular motion as in the present case. However, the excitations of
the accelerated systems in the two cases can be understood qualitatively in
the same way. The correlation functions $C_\pm$ gives a measure of vacuum
fluctuations of the electromagnetic field along the orbit $x(\tau)$, and
these fluctuations give rise to excitations in the detector when $C_+$
includes a spectral component which coincides with the excitation energy.

To see this more clearly let me first discuss the simplest case,
namely with a two level detector at rest. The transitions between the
levels are given by the same set of equations,
(\ref{gamma2}-\ref{D(tau)}), but now simply with
\be
x(\tau) = (\tau,0)
\ee
Since $D(\tau)$ is a linear combination of electromagnetic fields it can be
decomposed in the form
\be
D(\tau) = \int d^3k \sum_{r}[c_r(\vec{k})e^{-ikx}a_r(\vec{k}) +
d_r(\vec{k})e^{+ikx}a_r^\dagger (\vec{k})]
\ee
where $a_r(\vec{k})$ and $a_r^\dagger(\vec{k})$ are photon annihilation
and creation operators and $c_r(\vec{k})$ and $d_r(\vec{k})$ fourier
coefficients. According to eq.(\ref{gamma2}) only the positive
frequency parts of this operator are relevant for the transitions.
Positive frequency then is measured relative to the proper time along the
orbit $x(\tau)$. But with the detector at rest this coincides with
positive frquencies measured in the lab frame. And, as is well known, the
positive frequency part of lab frame fields only contains annihilation
operators. This is simply because all excitations in the lab frame have
positive energy. So the relevant component of $D(\tau)$ is
\be
\int_{-\infty}^{+\infty}d\tau e^{i\omega_0\tau}D(\tau) =
2\pi \int d\Omega_k\omega_0^2\sum_{r}c_r(\vec{k})
e^{i\vec{k} \cdot \vec{x}}a_r(\vec{k}) \ee
which only annihilates photons with the same energy as the energy
splitting of the two-level system. As a consequence of this the
probability for transitions up in energy is zero, since the $D$ operator acts
on the vacuum state. However, transitions down may be different from zero,
since in this case it is instead $D^\dagger$ which acts on the vacuum state.
Thus, the vacuum fluctuations only induce transitions to lower energies. This
clearly is related to energy conservation in the combined system of detector
and radiation field.

If the two-level system moves with constant velocity the picture is the
same, since the sign of the zero component of the photon momentum $k$ is
the same in all inertial frames. The only way to have a non-zero probability
for excitations to higher energies is to include other states than the
vacuum state in the one which the operator $D$ acts on. In particular
the probability is non-zero for states with temperature $T\neq 0$.

However, for accelerated motion this is no longer the case. $x(\tau)$ then
is no longer a linear function of $\tau$ and both functions
$e^{-ikx(\tau)}$ and $e^{+ikx(\tau)}$ will in general have positive
frequency parts in terms of the variable $\tau$. In addition, for the
electromagnetic field, there will be a $\tau$-dependent Lorentz
transformation connecting the fields in the co-moving frame with the lab
frame fields. The net effect is to introduce a mixing between the
positive and negative frequency parts, so that both the anihilation and
the creation parts of the operator $D$ will have positive frequency
components in terms of the time variable $\tau$. As a consequence of this
there will be in general non-vanishing probabilities for excitations both
up and down in energy for the accelerated system, even with the quantum
field in the vacuum state. 

For uniform linear acceleration along the z-axis, the accelerated path
$x(\tau)$ is described by 
\be t=\frac{a}{c} \sinh (\frac{a}{c}\tau),
\; z=\frac{a}{c} \cosh (\frac{a}{c}\tau), \; x=y=0,
\ee
The trajectory $x(\tau)$ in this case depends only on one free parameter,
which is the rest frame acceleration $a$. An interesting symmetry which is
present for this motion corresponds to a shift in the $\tau$-parameter in
the imaginary direction,
\be
x(\tau) = x(\tau + i \frac{2\pi c}{a})
\ee
This symmetry, together with general symmetries from field theory,
related to PCT-invariance\,\cite{Sewell82,Hughes85,Bell85}, gives a
simple relation between the correlation functions corresponding to
transitions up and down in energy, 
\be
C_+(\tau) = C_-(\tau-i\frac{2\pi c}{a})
\ee
This relation is similar to one which is present for correlation functions
at non-zero temperature, and it leads to a similar result for the ratio
between probabilities for transitions up and down,
\bea
\Gamma_+ &=& \int_{-\infty}^{+\infty} d\tau e^{i\omega_0\tau}C_-
(\tau-i\frac{2\pi c}{a}) \nonumber \\
&=& \int_{-\infty}^{+\infty} d\tau
e^{i\omega_0(\tau+i\frac{2\pi c}{a})}C_- (\tau) \nonumber  \\
 &=&\exp (-\frac{\hbar \omega_0}{a\hbar/2\pi c}) \Gamma_- 
\eea
If the ratio between the two transition probabilities now is interpreted
as a Boltzmann factor, then there is a simple linear relation between the
temperature associated with this factor and the acceleration $a$,
\be
kT_a = \frac{a\hbar}{2\pi c} \label{Unruh}
\ee
$T_a$ then is the Unruh temperature for the accelerated system and $k$
the Boltzmann constant. The derivation shows that the thermal property of
the excitation spectrum depends only on general properties of the quantum
fields and on special properties of the accelerated trajectory $x(\tau)$.
Details of the accelerated system is not important. 

In the case of electron polarization one may consider the question
whether linearly accelerated electrons could be used to detect the Unruh
effect. An additional magnetic field along the electron
path could provide the necessary splitting of the spin energy levels.  In
principle this should give a cleaner demonstration of the heating by
acceleration effect than the polarization effect for circulating electrons.
However, as discussed in ref.\,\cite{Bell83} the time needed to reach
equilibrium is far too long to make this effect relevant for the motion of
electrons in linear accelerators. For electrons in cyclic accelerators much
larger accelerations can be obtained and correspondingly much smaller time
constants for the approach to equilibrium.

\section{Unruh effect for circulating electrons}
Let me now consider the
electron polarization for circulating electrons from the point of view of the
Unruh effect. One main difference between circular motion and linear
acceleration is that the former depends on two independent parameters, which
we may take to be
$a$ and
$\gamma$. However, for ultrarelativistic electrons the quantum
fluctuations affecting the spin motion in the co-moving frame essentially
only depend on
$a$. If we disregard all complications related to the circular motion and
simply assume the temperature formula (\ref{Unruh}) to be valid, we find the
following. In the non-rotational C-frame the energy splitting of the spin
system is for $\gamma\gg1$,
\be
\Delta E = \hbar \omega_0 = g\frac{\hbar a}{2c}
\ee
This gives a Boltzmann factor
\be
\exp (-\frac{\Delta E}{kT_a}) = \exp [-\pi g] \label{Boltzmann}
\ee
and the polarization is 
\be
P(g) = \tanh [\pi g/2]
\ee
For the physical value $g=2$ this gives $P=0.996$ as compared with the
previously cited correct value $P=0.924$. 

It is of interest to notice the similarities and the differences in the
$g$-dependence of the correct function $P(g)$ and the function obtained from
the simple temperature formula. The two functions are displayed as curve A
and B in Fig.\,2. The main difference between these two curves is a shift
along the
$g$-axis. Such a shift may loosely be associated with an angular velocity
present in the system which couples to the spin. Thus, if we assume the spin
excitations to have a thermal form in the rotating O-frame rather than in the
C-frame, this would shift the polarization curve with two units along the
$g$-axis. The correct curve is located somewhere inbetween, and there seems
not to be a simple explanation for this. One should also note some important
differences in the details of the two curves.

In the temperature formula given above we have made the simple assumption
that the spin system can be considered as a thermally excited two-level
system, independent of the other degrees of freedom of the electron. This may
be a too simple model even if the Unruh temperature formula, in some
approximate meaning, should be valid. After all the fluctuations in the path
are important for the polarization effect, as I have previously pointed out.
In order to estimate the effect of these fluctuations, we may simply leave
out the two terms in (\ref{deltaomega}) associated with fluctuations in the
path and keep only the term which accounts for the direct coupling of the
magnetic moment to the quantum fluctuations of the magnetic field. The
resulting function is displayed as curve C in Fig.\,2. Now the shift
relative to the thermal curve has disappeared and the form is also quite
similar in the two cases. The new curve can in fact be approximated well by
a formula like (\ref{Boltzmann}), but with a temperature somewhat higher
than the Unruh temperature, $T_{eff}\approx 1.3 T_a$.  

\begin{figure}[t]
\centerline{\psfig{figure=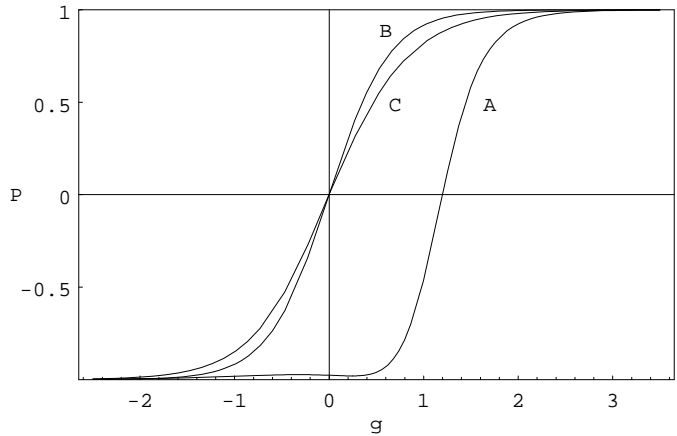,height=2in}}
\vspace{10pt}
\caption{{\em The electron polarization $P$ as a function of the gyromagnetic
factor $g$. The curve A is based on the detailed calculations. Curve B is
based on a two-level spin model with thermal excitations due to the Unruh
temperatur. Curve C is obtained from the detailed calculation (curve A)
when terms identified as due to orbital fluctuations are suppressed.}
\label{fig:P(g)}}
\end{figure}
Since the orbital fluctuations are important for the polarization
effect, a better comparison with the Unruh effect would be obtained by
treating the orbital motion together with the spin motion. However, as
already pointed out, the electron spin is not so important for the
orbital motion. As a final point I will therefore consider the vertical
motion without taking into account the effects of the spin. 

The (Heisenberg)
equation of motion for the vertical oscillations can be written as 
\be \ddot{z} + 2\Gamma \dot{z} + \Omega^2z = \frac{e}{m}E_{qz} 
\ee
In this equation $\Gamma=(e^2a^2)/(3mc^5)$ and only the most important
part of the radiation damping term has been kept. Making use of the fact
that the damping is small, $\Gamma \ll \Omega$, one can solve the equation 
to find an (approximate) expression for $z(\tau )$ in terms of the
quantum field $E_{qz}$. For the fluctuation in the $z$-coordinate one finds
the following expression 
\be 
<z^2> = \frac{1}{2\Gamma}(\frac{e}{m\Omega})^2
\int_{-\infty}^{+\infty} d\tau e^{-\Gamma |\tau|} \cos \Omega\tau
<E_{qz}(\tau/2)E_{qz}(-\tau/2)>  
\ee 
This shows that the fluctuations in the vertical direction are determined by
the correlation function of the z-component of the electric field along the
classical orbit. The vertical fluctuations in fact can be interpreted as
being due to the circular Unruh effect in a similar way as the polarization
effect. The mean energy associated with the fluctuations is for large
$\gamma$, \be <E>_{vert}\: = m\Omega^2<z^2>\: = \frac{13}{96}\sqrt{3}
\frac{a\hbar}{c} \ee It is proportional to the acceleration $a$, but with a
different prefactor as compared with the linear Unruh effect. It corresponds
to a somewhat higher temperature
\be
T_{eff} \approx 1.5 T_a
\ee
To linear order the excitation spectrum in this case in fact has a
thermal form, and there is no complication with rotating frames. So in
this respect the vertical orbit excitations give a simpler demonstration
of the Unruh heating in the circular case than the depolarization of the
electrons do. But the fluctuations are small and to measure them may be
much more difficult task.

\section{Concluding remarks}
The quantum effects for electrons in a storage ring have here been studied
within a simple idealized model for a cyclic accelerator. The electron spin
has been treated, in the co-moving accelerated frame of the classical orbit,
as a two-level system coupled linearly to the quantum fields and to the
orbital fluctuations. In this description the external magnetic field along
the orbit defines the (unperturbed)  spin levels of the electrons, and the
radiation field causes transitions between these two  levels. The radiation
field acts both directly on the spin, through the coupling to the magnetic
moment, and also indirectly, through the fluctuations it introduces in the
particle orbit. The equilibrium polarization calculated in this way agrees
with the classical results known from the literature. However, a resonance
between the spin and orbital fluctuations gives an effect which mainly is
depolarizing, but close to the resonance leads to a small increase in the
polarization. 

In a more realistic description of a cyclic accelerator there will be
several modifications of this picture. In the case of a strong focussing
machine the magnetic field will no longer be uniform along the orbit. The
unperturbed part of the spin Hamiltonian then will be time dependent in the
co-moving frame, and as a consequence of this the perturbations cannot be
described in terms of transitions between  stationary spin levels. There may
be other perturbations in the magnetic field that cause a coupling between
vertical and horizontal oscillations. Also non-linear effects may be
important. This will in general lead to a much richer structure of
spin-orbit resonances than in the idealized model where only one resonance
is present. All these effects certainly have to be taken into account when
one wants to model the spin behaviour in a real accelerator
\cite{Mane87,Kewisch89,Hand89}. Nevertheless, to understand the main aspects
of the quantum effects for the accelerated electrons, the simple idealized
model used here may be of interest.

As discussed in this paper, the expression for the spin flip probabilities
can be reduced to a form where they are determined only by vacuum correlation
functions of the electromagnetic fields along the classical orbit. Expressed
in this way there is a clear similarity between the polarization effect and
the Unruh effect for a linearly accelerated two-level system. But for
circular motion the correlation functions do not have a truly thermal form.
And for the circulating electrons there are complications due to the
rotations of frames along the orbit and due to coupling between the spin and
orbital fluctuations. However, when the effects of the orbital fluctuations
are supressed we obtain a polarization curve which is well approximated by
the curve obteined from a thermally excited two-level system.

A more careful comparison between the polarization effect and the Unruh
effect would mean to include the orbital motion in the description. Since the
fluctuations in the orbital motion can be treated as being independent of
the spin, it is meaningful to examine the vertical fluctuations separately.
Also these flutuations are determined by vacuum fluctuations of the
electromagnetic field along the classical orbit. The result for the vertical
fluctuations is that they do have a thermal excitation spectrum, but the
temperature is slightly higher than the Unruh temperature determined from the
acceleration alone.

\newpage


\end{document}